\documentclass[aps,prb,twocolumn]{revtex4}
\usepackage{graphicx,amsmath}

\begin{document}
\title{Distortion of the Stoner-Wohlfarth astroid by a spin-polarized current}
\author{Y. Henry}
\email{yves.henry@ipcms.u-strasbg.fr}\affiliation{Institut de Physique et
Chimie des Mat\'eriaux de Strasbourg, CNRS, Universit\'e de Strasbourg, B.P.
43, F-67037 Strasbourg Cedex 2, France}
\author{S. Mangin}
\affiliation{Institut Jean Lamour, Nancy Universit\'e, CNRS, B.P. 239, F-54506
Vandoeuvre Cedex, France}
\author{J. Cucchiara}
\affiliation{Institut Jean Lamour, Nancy Universit\'e, CNRS, B.P. 239, F-54506
Vandoeuvre Cedex, France}
\author{J.~A. Katine}
\affiliation{Hitachi GST San Jose Research Center, 3403 Yerba Buena Road, San
Jose, California 95135, USA}
\author{Eric~E. Fullerton}
\affiliation{Center for Magnetic Recording Research, University of
California-San Diego, La Jolla, California 92093-0401, USA}
%72.25.-b Spin polarized transport
%75.60.-d Domain effects, magnetization curves, and hysteresis
%75.75.+a Magnetic properties of nanostructures
%85.75.-d Magnetoelectronics; spintronics: devices exploiting...
%\date{\today}

\begin{abstract}
The Stoner-Wohlfarth astroid is a fundamental object in magnetism. It separates
regions of the magnetic field space with two stable magnetization equilibria
from those with only one stable equilibrium and it characterizes the
magnetization reversal of nano-magnets induced by applied magnetic fields. On
the other hand, it was recently demonstrated that transfer of spin angular
momentum from a spin-polarized current provides an alternative way of switching
the magnetization. Here, we examine the astroid of a nano-magnet with uniaxial
magnetic anisotropy under the combined influence of applied fields and
spin-transfer torques. We find that spin-transfer is most efficient at
modifying the astroid when the external field is applied along the easy-axis of
magnetization. On departing from this situation, a threshold current appears
below which spin-transfer becomes ineffective yielding a current-induced dip in
the astroid along the easy-axis direction. An extension of the Stoner-Wohlfarth
model is outlined which accounts for this phenomenon.
\end{abstract}

\maketitle

\section{Introduction\label{Introduction}}
The study of magnetization reversal has provoked continuous interest during the
last sixty years. Indeed, this topic has proven to be scientifically very
challenging, and one of the fundamental issues in magnetic data storage and
memory technologies. The difficulty to quantitatively understand magnetization
reversal, known as Brown's paradox,\cite{B45} results from complex domain
patterns that can form in magnetic materials. However, for nano-scale systems
the picture is greatly simplified, as initially described by Stoner and
Wohlfarth\,\cite{SW48} and N\'eel,\cite{N49} where the magnetic order parameter
is assumed to be uniform across the dimensions of the sample and can be
described as a single macro-spin. Fifty years were necessary to have the
technological environment needed to test the macro-spin model
experimentally.\cite{BWBB99} During that time, new techniques essential to
fabricate and characterized nanometer-size objects such as scanning probe
microscopies and nano-lithography were developed. It was then demonstrated
that, for some magnetic nanoparticules, the angular dependence of the switching
field can indeed follow the famous Stoner-Wohlfarth astroid curve which
separates regions of the cartesian magnetic field space with two stable
magnetic states from those with only one stable state.

Very few magnetic systems, however, possess the pure uniaxial magnetic
anisotropy required to closely obey the Stoner-Wohlfarth model and the locus in
the field space of all the magnetization switching fields of small magnets
seldom take the shape of an astroid, as defined mathematically. However, over
the years, the concept of Stoner-Wohlfarth astroid has been generalized to
arbitrary effective anisotropy.\cite{T00} The term is now commonly used to
refer to any critical surface that delimits the region of bistability of the
magnetization in the field space and is a fundamental property of magnetic
nano-materials.

More recently, it has been theoretically predicted\,\cite{S96,B96} and
experimentally evidenced\,\cite{KABM00} that injection of spin-polarized
electrical currents can induce magnetization reversal. This new approach for
magnetization reversal has generated considerable scientific interest and is
expected to play a major role in many emerging spintronic
technologies.\cite{KF08} Usually, transfer of angular momentum from a
spin-polarized current to the magnetization does not strongly modify the
positions of the magnetization equilibria.\cite{ST05} Its primary effect is to
produce an extrinsic damping which either reinforces or opposes the intrinsic
damping of the magnetization and modifies the stability of the equilibria
rendering, for example, an unstable equilibrium stable. Spin transfer is
expected to produce significant distortions of the astroid, making the region
of bistability expand in some parts of the magnetic field space and retract in
others. This distortion was considered theoretically by Sun\,\cite{S00} in the
case of an in-plane polarized current acting on a magnet with a combination of
uniaxial and easy-plane magnetic anisotropies.

In this paper, we examine how the Stoner-Wohlfarth astroid is indeed distorted
by spin-transfer in the simplest and most fundamental case of a nano-magnet
with uniaxial magnetic anisotropy submitted to a current polarized along the
easy-axis. As we do this, we reveal how the efficiency of the spin-transfer
torque with respect to magnetization switching varies in a rather
counter-intuitive manner with the orientation of the external magnetic field.

\section{Experiments\label{Experiments}}
The small magnet whose astroid is studied experimentally is the 3-nm thick
'free' element of a giant magnetoresistance (GMR) spin-valve patterned in the
shape of a vertical pillar with 100~nm $\times$ 200~nm hexagonal
cross-sectional shape. Both the free-element and the reference-element of the
device have perpendicular-to-plane magnetic anisotropy. This is achieved by
using [Co/Ni] and [Co/Pt]/[Co/Ni] multilayer stacks, respectively, as described
in details elsewhere.\cite{MRKC06} In comparison with the intrinsic
perpendicular-to-plane anisotropy of the extended Co/Ni multilayer, the shape
anisotropy introduced in the plane by patterning the film in a non-circular
pillar is negligible. Thus, the overall magnetic anisotropy of the free-element
is essentially \emph{uniaxial} and its astroid should be in first approximation
invariant upon rotation around the normal to the film plane $\hat{\mathbf{z}}$.

\begin{figure}
\includegraphics[width=8.5cm,trim=0 112 40 76,clip]{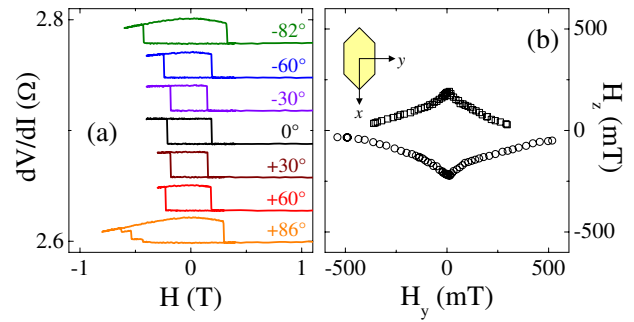}
\caption{(a) Examples of differential resistance versus magnetic field minor
loops of the free element in a 100~nm $\times$ 200~nm
[Co/Ni]/Cu/[Co/Ni]/[Co/Pt] nanopillar recorded at different angles of the
applied field with respect to the film normal. For clarity, the curves
corresponding to field angles other than $0^o$ are offset vertically. (b)
Experimentally determined two-dimensional cross-section of the zero-current
astroid of the free-element, in the ($H_y,H_z$) plane. Circles indicate a
transition from the parallel (low-resistance) state to the antiparallel
(high-resistance) state of the GMR device. Squares indicate the inverse
transition.} \label{Fig1}
\end{figure}

In practice, the astroid was built by determining the switching fields of the
free-element from differential resistance $dV/dI$ versus magnetic field minor
hysteresis loops [Fig.~\ref{Fig1}(a)] recorded at varying field angle
$\theta_H$ from $\hat{\mathbf{z}}$ ($-90^o\leq\theta_H\leq+90^o$), in the plane
defined by $\hat{\mathbf{z}}$ and the small axis of the hexagon
$\hat{\mathbf{y}}$ (see inset in Fig.~\ref{Fig1}(b)). The differential
resistance was measured using a Lakeshore Model 370 $AC$ resistance bridge with
an excitation current of $10~\mu$A RMS at 13.7~Hz. The $DC$ current provided by
a Keithley Model 2400 sourcemeter was injected in the sample using a home-made
bias-T interface. Before measuring each magnetoresistance loop, a positive
field in excess of 1~T was applied to ensure that both the free-element and the
reference-element would be initially magnetized positively, i.e. in the $z>0$
semi-space. The experiments have been performed several times and no strong
fluctuation of the switching fields, i.e. a stochastic behavior related to
thermal fluctuations, was observed. Moreover, no signature of
spin-transfer-induced steady precession states, i.e. non-hysteretic peaks or
dips in the differential resistance versus field loops,\cite{MRKC06} was found
with the $DC$ current values used.

In the off-axis geometries ($\theta_H\neq 0$), the magnetization vectors of the
two elements do not always remain strictly colinear to each other during field
cycling. This manifests in the curvature of both the decreasing-field branch
and, more obviously, the increasing-field branch of the measured GMR loops
[Fig.~\ref{Fig1}(a)]. However, for the sake of simplicity, we will still name
the low-resistance state and the high-resistance state of the nanopillar the
'parallel' (P) state and the 'antiparallel' (AP) state, respectively. Three
dimensional micromagnetic simulations performed with the OOMMF software
package\,\cite{OOMMF} indicate that the continuous change in the relative
orientation of the magnetization vectors that the curvature reveals is
primarily ascribable to the rotation of the magnetization of the free-layer. To
a good approximation, the magnetization of the reference-layer, harder
magnetically, remains fixed. An unambiguous determination of the switching
fields $H_{AP\rightarrow P}$ and $H_{P\rightarrow AP}$ was possible only for
field angles such that $|\theta_H|<+85^o$, where the reversals of the
free-element magnetization occur abruptly. For $85^o\leq|\theta_H|\leq 88^o$,
the GMR loops may contain several well-separated jumps (bottommost loop in
Fig.~\ref{Fig1}(a)) indicating that the reversal is non-uniform and sequential
either because it becomes dominated by pinning of domain walls on defects or,
equally likely, for intrinsic micromagnetic reasons such as the bifurcations
discussed in Ref.~\onlinecite{FTJW04}. For $|\theta_H|>88^o$, finally, sharp
discontinuities are no longer visible in the experimental curves. Hereafter, we
limit our discussion to angles less than 85$^o$.

\begin{figure*}
\includegraphics[width=18cm,trim=30 214 30 194,clip]{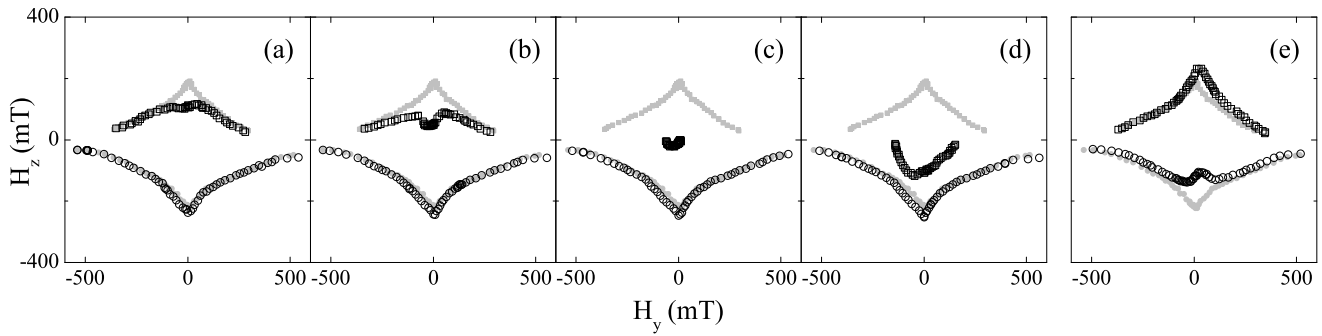}
\caption{Experimentally determined astroids of the free-element in a 100~nm
$\times$ 200~nm [Co/Ni]/Cu/[Co/Ni]/[Co/Pt] nanopillar for varying values of the
DC current density: (a) $j=+2\times 10^{11}$~A/m$^2$, (b) $+3\times
10^{11}$~A/m$^2$, (c) $+4\times 10^{11}$~A/m$^2$, (d) $+6\times
10^{11}$~A/m$^2$, and (e), $-6\times 10^{11}$~A/m$^2$. The zero-current astroid
is also shown for comparison (solid grey symbols).} \label{Fig2}
\end{figure*}

The zero-current astroid of the free-element is shown in Fig.~\ref{Fig1}(b). It
is qualitatively similar to the astroid of an isolated single-domain particle.
However, two differences may be pointed out. i) First, the experimental astroid
is comparatively flatter: The switching field close to the hard-axis is
markedly larger than that along the easy-axis. This is an indication that the
lateral dimensions of the free-element are too large for truly coherent
rotation of the magnetization to occur throughout the entire range of angles
explored. Indeed, in the literature, demonstrations of coherent rotation
behavior exist only for smaller magnetic objects.\cite{BWBB99} Full
micromagnetic simulations reproduce this flattening and show that, in the
parallel state, an $S$-like non-uniform magnetization pattern forms in the
free-layer at large field angles. ii) Second, unlike the square astroid of a
Stoner particle, the astroid of the free-element shows a significant asymmetry
with respect to the hard-axis of magnetization. This is due to the sizable
dipolar coupling that exists between the two magnetic elements of the pillar
only separated by a 4~nm thick copper spacer. The reference-element magnetized
along $\hat{\mathbf{z}}$ produces an average stray field of about 25~mT which
tends to maintain the magnetization of the free-element oriented upwards. This
results in a displacement of the astroid towards negative $H_z$, quite as for
an exchange-biased ferromagnetic film.\cite{SSKJ03}

We now extend these studies to include the contribution of spin-polarized
currents to reversal. To avoid damage of the nanopillar during measurements, we
have had to limit our investigations to moderate current densities ($|j|\leq
6\times 10^{11}$~A/m$^2$). Moreover, the system studied behaves in such a
way\,\cite{MRKC06} that positive currents which become spin-polarized by
transmission through the reference-layer are much more efficient at modifying
the switching fields of the free-element (especially $H_{AP\rightarrow P}$)
than negative ones polarized on reflection from the reference-layer. Therefore,
we are mainly presenting here results obtained with positive currents.

Positive currents favor an alignment of the free-element magnetization parallel
to that of the reference-element. As may be seen in Fig.~\ref{Fig2}(a-d), the
spin-transfer effect they generate modifies strongly the upper half of the
astroid, that is, the part of the critical curve corresponding to the switching
from the antiparallel state to the parallel state ($H_{AP\rightarrow P}$). As
spin-transfer starts to operate, the easy-axis cusp present for zero $DC$
current [Fig.~\ref{Fig1}(b)] disappears. Instead, a dip forms along the
easy-axis direction which gets deeper and broader as the current increases. At
first, the large angle parts of the $H_{AP\rightarrow P}$ branch of the astroid
remains relatively unchanged [Fig.~\ref{Fig2}(a,b)]. At larger currents, the
entire branch which can be probed with radial fields (constant $\theta_H$) is
affected and takes a semi-circular shape [Fig.~\ref{Fig2}(c,d)]. In contrast,
the lower half of the astroid corresponding to the reversal from the parallel
state to the antiparallel state ($H_{P\rightarrow AP}$) is not strongly
affected under positive current. The domain of stability of the parallel state
only slightly expands in the direction of negative $H_z$ and the easy-axis cusp
remains visible.

For negative currents [Fig.~\ref{Fig2}(e)], a qualitatively symmetrical
behavior is observed. On the side of the $H_{P\rightarrow AP}$ branch, the
astroid shrinks back as a dip forms along the easy-axis direction. On the side
of the $H_{AP\rightarrow P}$ branch, the astroid noticeably expands towards
positive $H_z$. Quantitative differences exist though (e.g., the size of the
dip) between astroids obtained for currents of the same magnitude but opposite
polarities (compare Figs.~\ref{Fig2}(d) and \ref{Fig2}(e)). These are
straightforwardly related to the difference in the spin-transfer efficiency
between the two current directions. The fact that upon injection of currents of
the two polarities, large parts of the critical curve, and sometimes one half
of it, remain virtually unchanged is a strong indication that the heat and
Oersted field generated by passing the current through the device are not
sufficient to modify the astroid and cannot be evoked to account for the
observed distortions.\cite{JWTM01}

In the remainder of the paper, we will mostly concentrate on those parts of the
astroid where spin-transfer gives rise to a reduction of the switching field,
that is, in those regions where the dip forms, which are the most relevant to
technological applications. Theoretical investigations\,\cite{BJZ04} have shown
that in the on-axis geometry ($\theta_H=0$), linear relations should exist
between the values of $H_{P\rightarrow AP}$ and $H_{AP\rightarrow P}$ and the
magnitude of the $DC$ current injected, in partial agreement with experimental
findings.\cite{MRKC06} The present results reveal that such is not the case if
the field is applied at a large angle away from the easy-axis. More
specifically, for every non-zero field angle, a threshold current $j_{min}$
exists below which spin-transfer does not affect magnetization reversal. This
minimum current increases with increasing $\theta_H$. As a consequence for
\emph{moderate} current values, there exist a field angle $\theta_H^{max}$
above which spin-transfer becomes ineffective. These results are rather
counterintuitive. Indeed, as $\theta_H$ increases, so does the relative angle
between the magnetization vectors of the reference and free layers and,
consequently, so does the spin-transfer torque. Naively, one might therefore
expect an enhanced efficiency of spin-transfer at large field angle.

\begin{figure*}
\includegraphics[width=18cm,trim=30 214 30 172,clip]{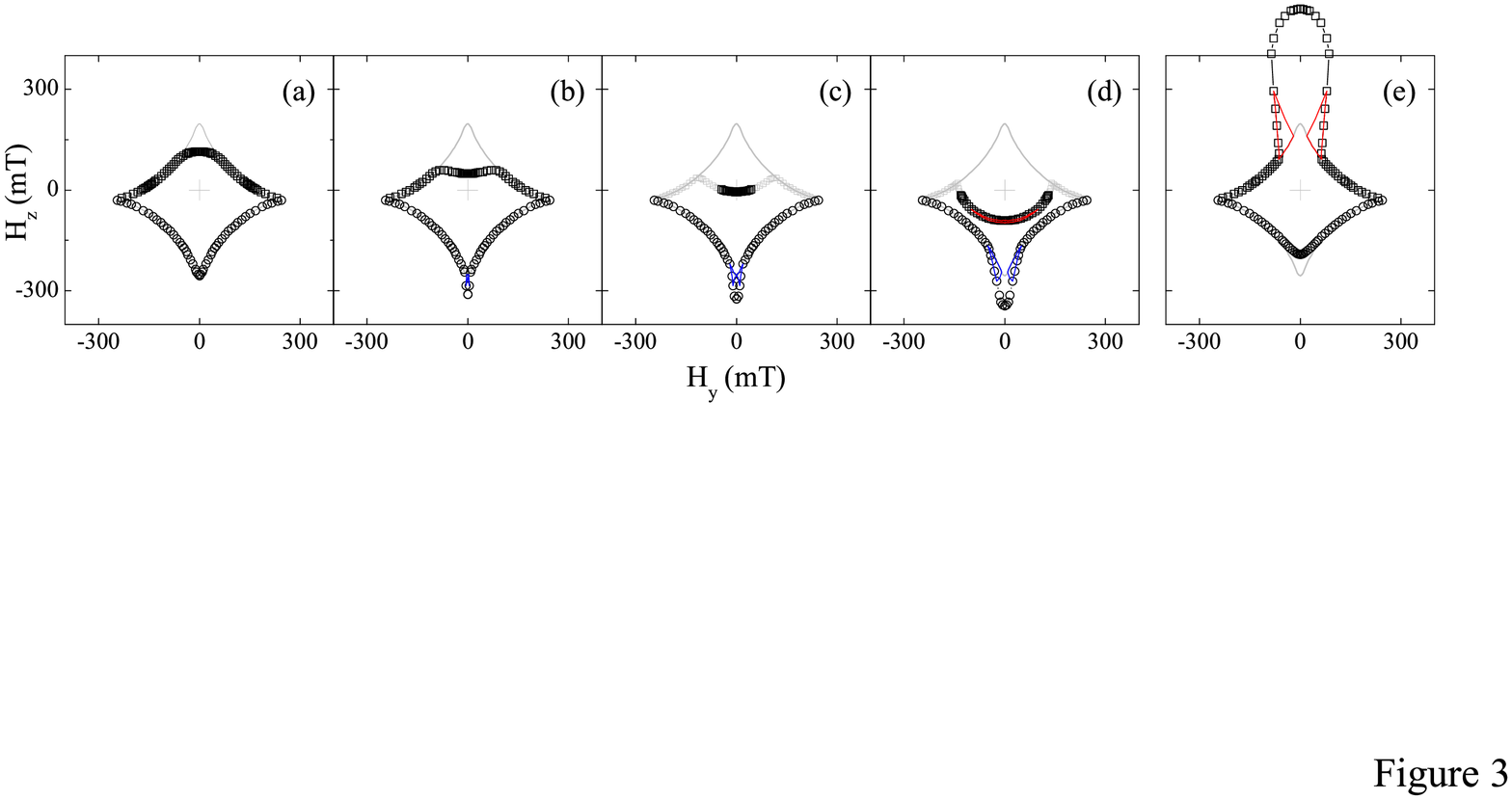}
\caption{Calculated astroids of the free-element. The values of the DC current
density used in the simulations are the same as experimentally
(Fig.~\ref{Fig2}): (a) $j=+2\times 10^{11}$~A/m$^2$, (b) $+3\times
10^{11}$~A/m$^2$, (c) $+4\times 10^{11}$~A/m$^2$, (d) $+6\times
10^{11}$~A/m$^2$, and (e) $-6\times 10^{11}$~A/m$^2$. The large black symbols
indicate the portions of the astroid accessible upon sweeping the magnetic
field at constant $\theta_H$. The smaller grey symbols are the complementary
parts determined from field sweeps at constant $H_y$. The red (respectively
blue) lines delineate regions of the field  space where the stationary
antiparallel state (respectively parallel state) is replaced by a steady
precession state with large negative (respectively positive) $m_z$. The astroid
computed for $j=0$ is also shown for comparison (grey line).} \label{Fig3}
\end{figure*}

\section{Numerical simulations\label{Simulations}}
To gain a deeper understanding of the effect of spin-transfer on the astroid,
we carried out numerical simulations in a macro-spin approach where the
magnetization of the reference layer is supposed to be fixed. By comparing the
experimental results with those from such a model, our goal is to unravel what
is intrinsic to the physics of spin-transfer and what is possibly due to sample
imperfections, higher-order magnetic anisotropies or deviations from a uniform
magnetization distribution.

In order to include a description of the dipolar interactions in the system as
realistic as possible under the assumption of uniform magnetization, the two
magnetic elements of the nanopillar were assumed to be identical, 200~nm long,
100~nm wide, 3~nm thick parallelepipeds, separated vertically by 4~nm.
Analytical results from Newell \textit{et al.}\cite{NWD93} were used to
calculate two tensors; first, the self demagnetizing tensor $N_D$, which
relates the demagnetization field $\mathbf{H}_i^{D}$ inside parallelepiped $i$
to its magnetization $\mathbf{M}_i$ through
$\mathbf{H}_i^{D}=-N_D\cdot\mathbf{M}_i$
\begin{equation}
N_D=\left|\begin{array}{ccc}0.0213&0&0\\0&0.0437&0\\0&0&0.9350
\end{array}\right| \label{N_D}
\end{equation}
and second, the so-called mutual demagnetizing tensor $N_M$ which allows one to
give a simple expression for the stray field $\mathbf{H}_i^{dip}$ generated by
element $j$ (magnetization $\mathbf{M}_j$) and experienced by element $i$,
$\mathbf{H}_i^{dip}=-N_M\cdot\mathbf{M}_j$
\begin{equation}
N_M=\left|\begin{array}{ccc}0.0105&0&0\\0&0.0217&0\\0&0&-0.0322
\end{array}\right| \label{N_M}
\end{equation}
Both the demagnetizing field $\mathbf{H}^{D}$ and the stray field from the
reference-element $\mathbf{H}^{dip}$ were included in the effective field
acting the magnetization of the free-element. For the latter element, we
assumed a saturation magnetization of $M_S=650$~kA/m, a perpendicular-to-plane
magnetic anisotropy constant $K_{\perp}=3\times 10^5$~J/m$^3$ and a damping
parameter $\alpha=0.01$, whereas for the reference-element we used
$M_S^{ref}=500$~kA/m (Ref.~\onlinecite{MRKC06}).

The switching fields were extracted from magnetization versus field loops
calculated by solving numerically, with the fourth-order Runge-Kutta algorithm,
a modified Landau-Lifschitz-Gilbert (LLG) equation including an additional
spin-transfer torque acting on the magnetization $\mathbf{M}$ of the form
proposed by Slonczewski\,\cite{S96,S02}, that is,
$-\beta(\theta)\,j/M_S\,[\mathbf{M}\times(\mathbf{M}\times\hat{\mathbf{p}})]$,
with the direction of spin polarization $\hat{\mathbf{p}}=\hat{\mathbf{z}}$.
The variation of the spin-transfer efficiency function $\beta$ with the
relative angle $\theta$ between $\hat{\mathbf{p}}$ and the magnetization
direction $\hat{\mathbf{m}}=\mathbf{M}/M_S$ assumed in these calculations is of
the same form as that found in Ref.~\onlinecite{SM06}
\begin{equation}
\beta(\hat{\mathbf{m}}\cdot\hat{\mathbf{p}})=\frac{\gamma\hbar}{2d\mu_0M_Se}
\left[\frac{q_+}{b_0+b_1\,(\hat{\mathbf{m}}\cdot\hat{\mathbf{p}})}+
\frac{q_-}{b_0-b_1\,(\hat{\mathbf{m}}\cdot\hat{\mathbf{p}})}\right]
\label{BetaSM}
\end{equation}
where $d$ is the thickness of the free-layer, $\gamma>0$ is the gyromagnetic
ratio, and $e>0$ is the absolute value of the electron charge. The parameters
$q_+=1/60$, $q_-=-1/600$, $b_0=1$, and $b_1=1/2$ were chosen in the following
way. $b_0$ was arbitrarily set to 1. The ratio $b_0/b_1$ was set to 2 so as to
obtain a significant asymmetry of the slope of the spin-transfer-torque between
angles $\theta$ close to $0^o$ and those close to $180^o$. The ratio $q_+/q_-$
was set to $-10$ to account for the asymmetry of the GMR device architecture.
Finally, the ratio $b_0/q_+=60$ was adjusted so that values of the current
density used in the simulations would approximately match those in the
experiments.

\begin{figure}
\includegraphics[width=8.5cm,trim=-15 76 -15 54,clip]{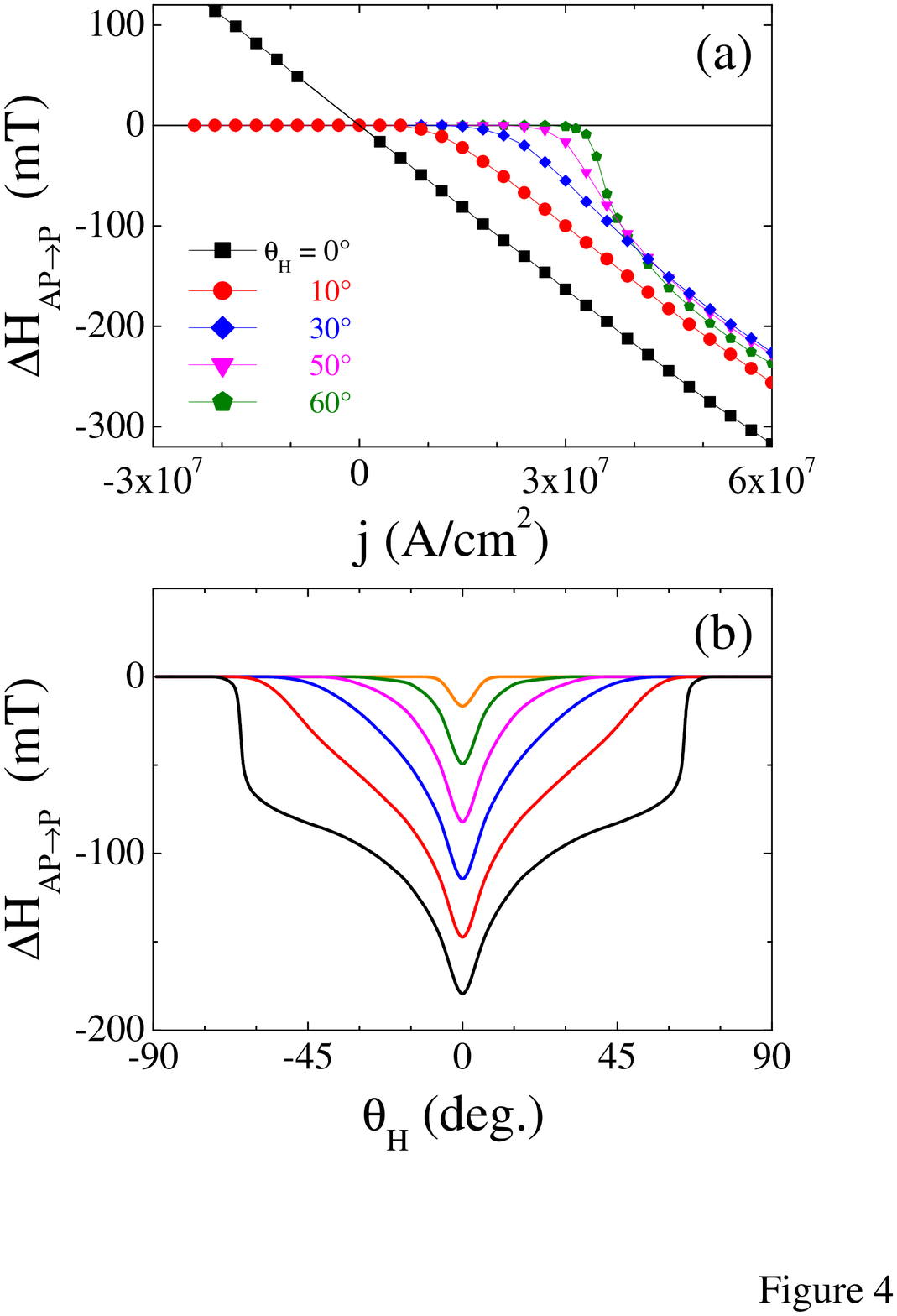}
\caption{Calculated change in the switching field for the transition from the
AP state to the P state. (a)  $\Delta H_{AP\rightarrow
P}(j,\theta_H)=H_{AP\rightarrow P}(j,\theta_H)-H_{AP\rightarrow P}(0,\theta_H)$
versus current density $j$ for various values of the applied field angle
$\theta_H$. (b) $\Delta H_{AP\rightarrow P}$ versus field angle $\theta_H$ for
equally spaced values of the current density $j$ ranging from $+0.6\times
10^{11}$~A/m$^2$ (topmost, orange curve) to $+3.6\times 10^{11}$~A/m$^2$
(bottommost, black curve).} \label{Fig4}
\end{figure}

No effort was made to adjust these parameters further in order to get the best
possible agreement with experimental data and reach a quantitative agreement,
should this be possible. Yet, as may be seen in Fig.~\ref{Fig3}, most of the
important features of the spin-transfer-distorted astroid discussed before are
accounted for by this approach. In particular, the existence of a threshold
current when $\theta_H\neq 0$ is confirmed. This is even more clearly seen in
Fig.~\ref{Fig4}(a) which plots the change in the switching field $\Delta
H_{AP\rightarrow P}(j,\theta_H)=H_{AP\rightarrow
P}(j,\theta_H)-H_{AP\rightarrow P}(0,\theta_H)$ as a function of the current
density, for various field orientations. Also obvious from Fig.~\ref{Fig4}(a)
is that the larger $\theta_H$ the less linear the variation of the switching
field with $j$, beyond $j_{min}$. Figure~\ref{Fig4}(b) illustrates the loss of
efficiency of spin-transfer beyond an angle $\theta_H^{max}$ which increases
with increasing $j$.

The absence of a large bubble-shaped extension of the astroid along the
$H_z$-axis, for negative current, is the only important point of disagreement
between modeling [Fig.~\ref{Fig3}(e)] and experiments [Fig.~\ref{Fig2}(e)].
Numerically, the extent of this protusion is found to be strongly reduced if
the easy-axis of magnetization makes an angle of a few degrees with respect to
the magnetic field plane. A slight distribution in the orientation of the
easy-axis in the free layer of the studied device is thus a possible
explanation for the discrepancy. We note however that the choice of the spin
transfer efficiency function $\beta$ affects also the size of this bubble.

\begin{figure}
\includegraphics[width=8.5cm,trim=70 336 70 264,clip]{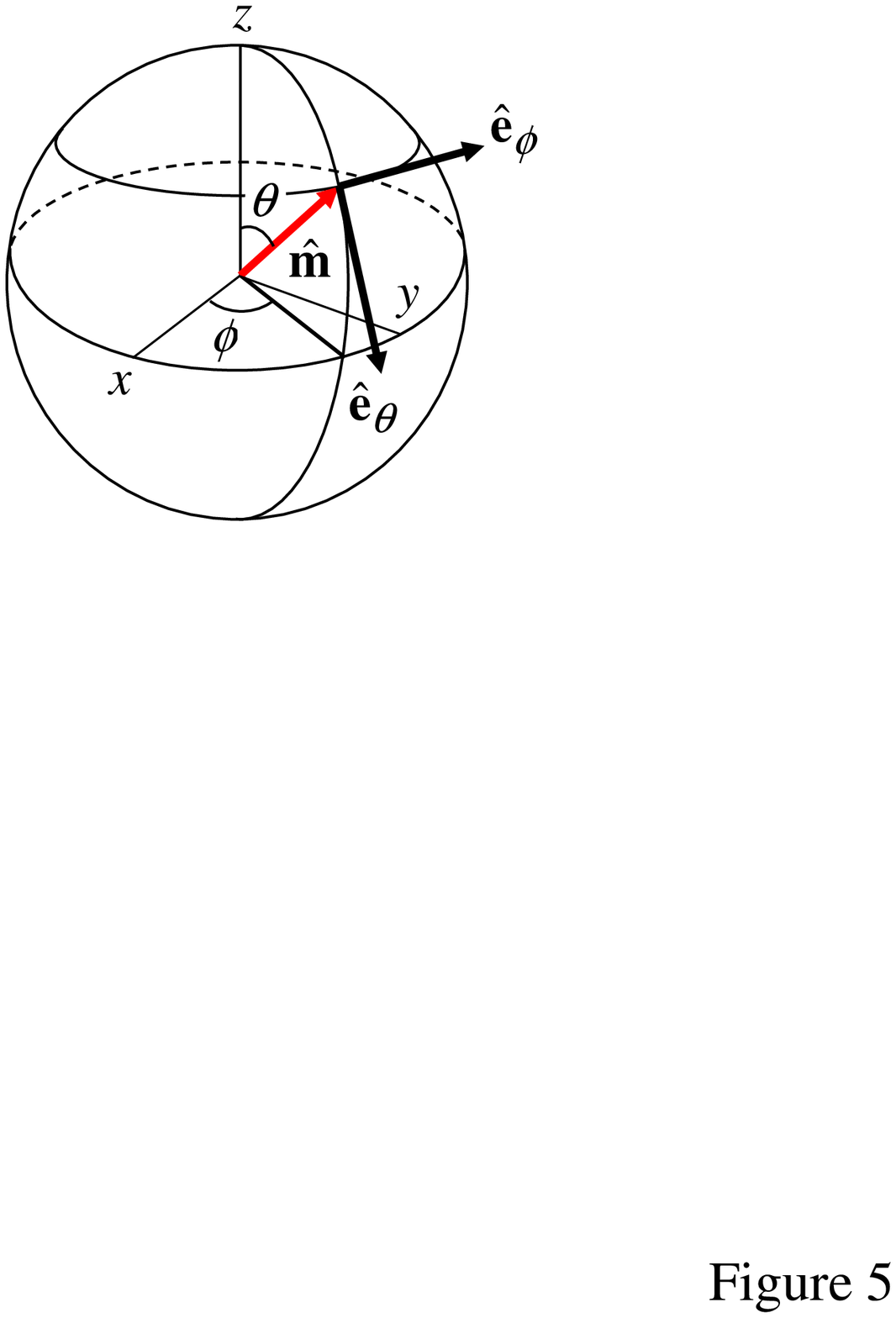}
\caption{(a) Polar and azimuthal angles $\theta$ and $\phi$, and vector basis
$\{\hat{\mathbf{m}},\hat{\mathbf{e}}_{\theta}, \hat{\mathbf{e}}_{\phi}\}$.}
\label{Fig5}
\end{figure}

\section{Analytical modeling\label{Modeling}}
Interestingly, a simple extension of the Stoner-Wohlfarth (SW)
model\,\cite{SW48} is sufficient to capture most features of the
current-distorted astroid. Let us consider a uniformly magnetized nano-magnet
with saturation magnetization $M_S$ and uniaxial magnetic anisotropy of axis
$\hat{\mathbf{z}}$ and constant $K$ ($H_K=2K/\mu_0M_S$). Its magnetization
$\mathbf{M}=M_S\,\hat{\mathbf{m}}$ is described by a polar angle $\theta$
measured from $\hat{\mathbf{z}}$ and an azimuthal angle $\phi$ measured from
$\hat{\mathbf{x}}$ [Fig.~\ref{Fig5}]. As is the case experimentally, let us
assume that the external field is applied in the $yz$ plane. In the orthonormal
direct basis formed by the three vectors $\hat{\mathbf{m}}$,
$\hat{\mathbf{e}}_{\theta}=\partial\hat{\mathbf{m}}/\partial\theta$,
$\hat{\mathbf{e}}_{\phi}=(1/sin\theta)\partial\hat{\mathbf{m}}/\partial\phi$,
the modified LLG equation that governs the dynamics of $\hat{\mathbf{m}}$ may
be written as
\begin{equation}
\frac{d\hat{\mathbf{m}}}{dt}=-\gamma(\hat{\mathbf{m}}\times\mathbf{H}_{eff}^*)
+\alpha(\hat{\mathbf{m}}\times\frac{d\hat{\mathbf{m}}}{dt})
\label{LLG}
\end{equation}
where $\alpha$ is Gilbert's damping constant. The total effective field acting
on $\mathbf{M}$
\begin{equation}
\mathbf{H}_{eff}^*=\mathbf{H}+H_K(\hat{\mathbf{m}}\cdot\hat{\mathbf{z}})
\hat{\mathbf{z}}+j\,\frac{\beta(\hat{\mathbf{m}}\cdot\hat{\mathbf{z}})}
{\gamma}(\hat{\mathbf{m}}\times\hat{\mathbf{p}}) \label{EffField*}
\end{equation}
is the sum of the usual effective field
\begin{equation}
\mathbf{H}_{eff}=-\frac{1}{\mu_0M_S}\frac{\partial E}{\partial\hat{\mathbf{m}}}
\label{EffField}
\end{equation}
which derives from the magnetic energy
\begin{equation}
E=-\mu_0\mathbf{M}\cdot\mathbf{H}-K(\hat{\mathbf{m}}\cdot\hat{\mathbf{z}})^2
\label{Energy}
\end{equation}
and of the spin-torque field
\begin{equation}
\mathbf{H}_{ST}=j\,\frac{\beta}{\gamma}\,
(\hat{\mathbf{m}}\times\hat{\mathbf{z}})=-j\,\frac{\beta}{\gamma}\sin\theta\,
\hat{\mathbf{e}}_{\phi}. \label{STField}
\end{equation}

At equilibrium, $\hat{\mathbf{m}}$ is necessarily parallel to
$\mathbf{H}_{eff}^*$. Therefore, the equilibrium conditions are
\begin{subequations}
\begin{eqnarray}
(\mathbf{H}_{eff}^*\cdot\hat{\mathbf{e}}_{\theta})_{_0}&=&0\label{EqCond1}\\
(\mathbf{H}_{eff}^*\cdot\hat{\mathbf{e}}_{\phi})_{_0}&=&0\label{EqCond2}
\end{eqnarray}
\end{subequations}
where the subscript "0" denotes equilibrium.

The consequence of introducing spin-transfer is twofold. (i) First, since
$\mathbf{H}_{eff}^*$ may have a component along $\hat{\mathbf{e}}_{\phi}$
[Eqs.~\ref{EffField*},\ref{STField}], $\hat{\mathbf{m}}$ does not always lie in
the plane defined by the applied field $\mathbf{H}$ and the easy-axis
$\hat{\mathbf{z}}$, at equilibrium. Therefore, the problem becomes
three-dimensional, in general. In the limit of small currents (and/or small
field angles), however, deviations of $\hat{\mathbf{m}}$ from the
($\hat{\mathbf{z}},\mathbf{H}$) plane remain small and, in first order
approximation, the problem can still be treated as if it were two-dimensional
($\phi=\pi/2$). (ii) Second, the stability of the equilibria can no longer be
determined from free energy considerations only and a new stability criterion
must be derived. This is possible analytically in the small current limit.
Indeed, performing a linear stability analysis around an equilibrium position
$\theta_0$, i.e. relating stability to the gradient of the total torque along
$\hat{\mathbf{e}}_{\theta}$, leads to the criterion
\begin{equation}
\left.\frac{\partial}{\partial\theta}[\alpha(\mathbf{H}_{eff}\cdot
\hat{\mathbf{e}}_{\theta})-j\,\frac{\beta(\theta)}{\gamma}\,\sin\theta]
\right|_{\theta=\theta_0}\leq 0 \label{StabCrit}
\end{equation}
The above criterion is written so as to reveal the competition between two
terms related to the torques produced by the effective field $\mathbf{H}_{eff}$
and the spin-torque field $\mathbf{H}_{ST}$, respectively. More practically,
equation~\ref{StabCrit} shows also that stability in the off-axis case
($\theta_H\neq 0$, $\theta_0\neq 0[\pi]$) is influenced not only by the
spin-transfer efficiency function $\beta(\theta)$ but also by its derivative
$\partial\beta/\partial\theta$, a point first put forward by Smith \textit{et
al.}\cite{SKCC05}

\begin{figure}
\includegraphics[width=8.5cm,trim=10 280 10 234,clip]{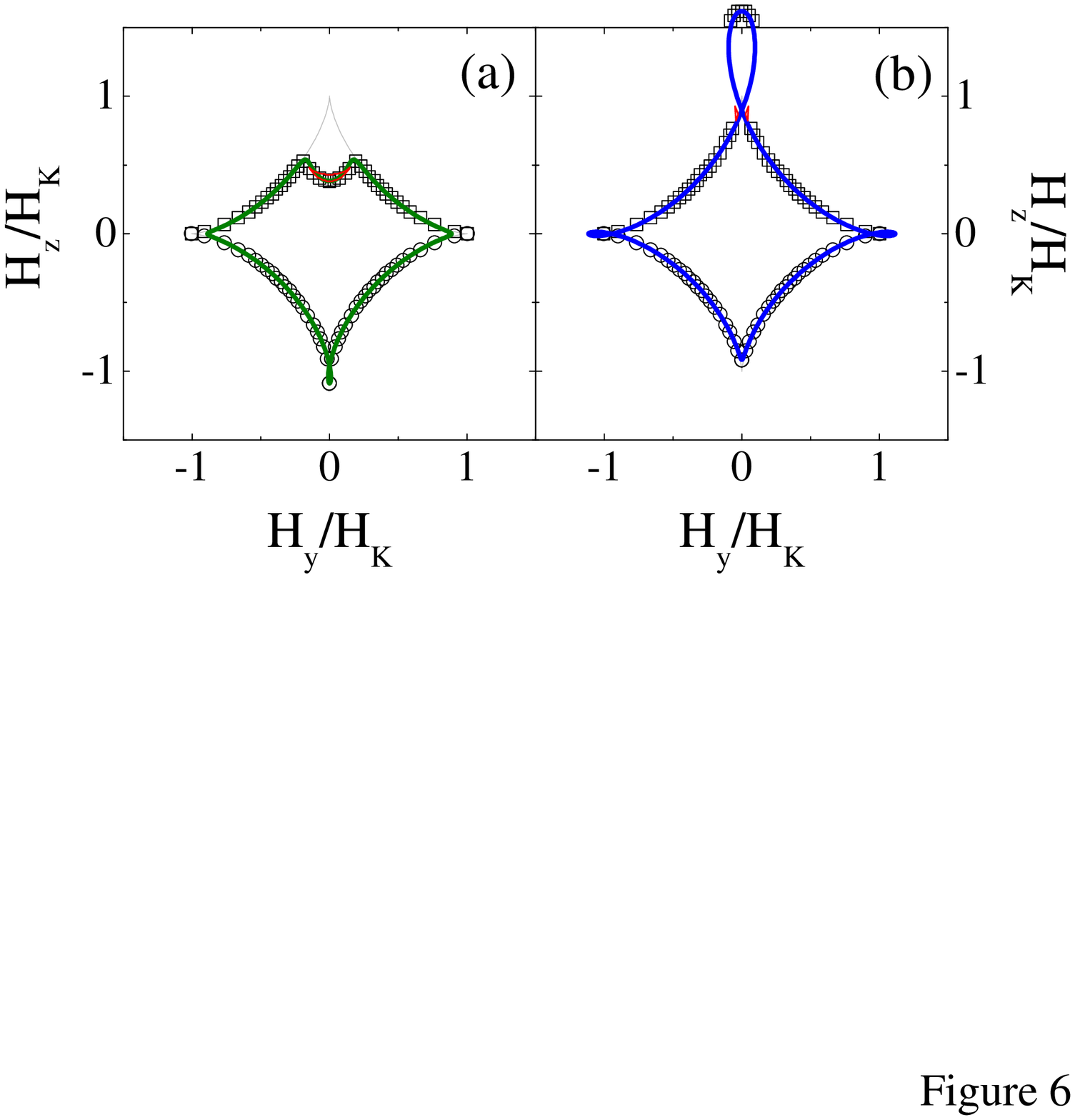}
\caption{Comparison of astroids drawn from the extended Stoner-Wohlfarth model,
Eqs.~\ref{ParamEq1} and \ref{ParamEq2}, (thick lines) and deduced from
macro-spin simulations (symbols) in the case of a small positive current,
$j=+3\times 10^{9}$~A/m$^2$ (a) and a small negative current, $j=-3\times
10^{9}$~A/m$^2$ (b). As in Fig.~\ref{Fig3}, the thin red lines delineate
regions of the field space where the antiparallel state is replaced by a steady
precession state with large negative $m_z$.} \label{Fig6}
\end{figure}

By combining the equilibrium condition, Eq.~\ref{EqCond1}, and the stability
criterion, Eq.~\ref{StabCrit}, we can derive the following parametric equations
of the astroid under small current
\begin{equation}
\left\{\begin{array}{ll}H_y=&H_K\sin^3\theta_0-j\,\sin\theta_0\,C(\theta_0)\\
H_z=-&H_K\cos^3\theta_0-j\,\cos\theta_0\,C(\theta_0) \end{array}\right.
\label{ParamEq1}
\end{equation}
with
\begin{equation}
C(\theta_0)=\frac{1}{\alpha\gamma}\left.\frac{\partial(\beta\sin\theta)}
{\partial\theta}\right|_{\theta=\theta_0} \label{ParamEq2}
\end{equation}
where the correction terms due to spin-transfer appear most clearly. As may be
seen in Fig.~\ref{Fig6}(a,b), predictions from this analytical model regarding
the stability of the equilibria closely agree with results from computer
simulations using the same assumptions. Of course, the model is unable to
predict the existence of dynamic states and regions of the field space where
two magnetic states exist but only one is stationary (as indicated by numerical
simulations) are not comprised in the astroid thus defined. In the extended SW
model, the negative counterpart of the dip is a kind of protruding bubble
attached to the rest of the astroid through a crunode (double point,
Fig.~\ref{Fig6}(b)).

For the numerical application of Eq.~\ref{ParamEq1} and the comparative
macro-spin simulations shown in Fig.~\ref{Fig6}, the overall magnetic
anisotropy constant $K$ was set to $K_{\perp}-\mu_0 M_S/2$, with the values of
$K_{\perp}$ and $M_S$ given in section~\ref{Simulations}. Moreover, we used an
expression of $\beta$ in agreement with Slonczewski's original
proposition\,\cite{S96}
\begin{equation}
\beta(\hat{\mathbf{m}}\cdot\hat{\mathbf{p}})=\frac{\gamma\hbar}{2d\mu_0M_Se}
\left[\frac{1}{f(P)(3+\hat{\mathbf{m}}\cdot\hat{\mathbf{p}})-4}\right]
\label{BetaSlonc1}
\end{equation}
with
\begin{equation}
f(P)=\frac{(1+P)^3}{4P^{3/2}} \label{BetaSlonc1}
\end{equation}
and a degree of spin polarisation of the electrons coming out of the reference
layer arbitrarily set to $P=50~\%$. For demonstration purposes, this function
was deliberately chosen different from the one used in the numerical
simulations of Fig.~\ref{Fig3} [Eq.~\ref{BetaSM}] although both are monotonous,
increasing functions of $\theta$ on the $[0,\pi]$ interval. Yet, the two
functions which were previously considered in the literature lead to
qualitatively identical modifications of the astroid shape and to the existence
of an angle $\theta_H^{max}$ above which the switching field is virtually
unchanged. This demonstrates once more the robustness of this feature.

\section{Discussion}
The magnitude of the effective field acting on the magnetization at the very
beginning of reversal is the key parameter to consider in order to understand
the effect of spin-transfer on the astroid. The reason for that is that
$H_{eff}$ is a measure of the intrinsic damping. Indeed, the larger $H_{eff}$,
the faster the precession of $\mathbf{M}$ around $\mathbf{H}_{eff}$, in the
event of an excursion away from the equilibrium, and the stronger the viscous
damping which drives $\mathbf{M}$ back to equilibrium. From Eqs.~\ref{EqCond1}
and \ref{StabCrit}, one can readily derive that along the zero-current astroid
\begin{equation}
H_{eff}(\theta_0)=H_K\sin^2\theta_0. \label{HeffonAstroid}
\end{equation}

The on-axis geometry is extremely particular. In this geometry, the effective
field and the intrinsic damping vanish at the zero-current switching fields
[Eq.~\ref{HeffonAstroid}]. This is why, close to these field values, any small
amount of extrinsic damping brought in by spin-transfer affects the
magnetization reversal. Furthermore, the orientation of the magnetization being
independent of the applied field ($\theta_0=0$ or $\pi$), $H_{eff}$ varies
linearly with $H$ and the extrinsic damping due to spin-transfer is just
proportional to $j$. This leads to the observed linear relation between the
switching field and the injected current [Fig.~\ref{Fig4}(a)].

In the off-axis geometries, the situation is qualitatively different. Indeed,
the effective field always retains a sizable magnitude and the magnetization
experiences a finite intrinsic (positive) damping at reversal. To induce an
early switching of the magnetization, e.g. $H_{AP\rightarrow
P}(j)<H_{AP\rightarrow P}(0)$, spin transfer must produce enough negative
damping to overcome this finite intrinsic damping. $j_{min}$ is the smallest
current density which realizes this. If $j$ is less than $j_{min}$ then,
irrespective of $j$, reversal occurs when the equilibrium loses its local
stability, i.e. upon crossing (exiting) the zero-current astroid, as for $j=0$.
The mathematical complexity of the $\theta_H\neq 0$ case, which is largely due
to the fact that the orientation of the magnetization changes continuously with
$H$, makes it difficult to derive an analytical expression for the dependence
of $j_{min}$ on $\theta_H$. To explain why $j_{min}$ increases with $\theta_H$
or, equivalently, why a moderate current density generates enough negative
damping only up to a given field angle $\theta_H^{max}$ one has to invoke the
fact that the intrinsic (positive) damping increases faster with $\theta_H$
than the extrinsic (negative) one.

\section{Conclusion}
In summary, we have investigated the effect of spin-transfer on the
Stoner-Wohlfarth astroid of a small magnet with uniaxial magnetic anisotropy.
The distortions observed experimentally are well accounted for qualitatively by
both macro-spin numerical simulations and a simple extension of the SW
analytical model. Evidence has been given that spin-transfer is more efficient
at modifying the switching field in geometries close to the so-called axial
geometry\,\cite{BJZ04} where the external field is applied along the easy-axis
and spin polarisation direction. On departing form this situation, a threshold
current appears below which spin-transfer is ineffective, the larger the field
angle the larger this current.

Our results have implications for solid-state device applications. For
zero-current, the SW model predicts that a two-fold reduction of the switching
field can be achieved by applying the external field at an angle of $45^o$ from
the anisotropy axis, a strategy often used in magnetic memory cells. What our
results suggest is that combining such a strategy with spin-transfer will not
necessarily help in reducing the switching field further. For the injection of
current to be efficient, its density will have to exceed the threshold value
$j_{min}$. We note finally that the physics which describes the distortions of
the Stoner-Wohlfarth astroid under the influence of spin-polarized currents is
similar to that which explains spin-transfer-induced instabilities of the
sensing layer in GMR read heads of hard disk drives.\cite{SKCC05}

\begin{acknowledgments}
The authors thank M. Bailleul, E. Bonet, J. Grollier, A.~D. Kent, P. Panissod,
D. Ravelosona, J. Sun, A. Thiaville, C. Thirion, and I. Tudosa for insightful
discussions, as well as O. Bengone for assistance in symbolic computations.
\end{acknowledgments}

\end{document}